\documentclass[12pt]{iopart} 
\usepackage{epsfig}
\usepackage{iopams}
\usepackage{epsfig}
\usepackage{graphicx}
\usepackage{graphics}
\usepackage{subfigure}
\usepackage{verbatim}
\usepackage{color}
\newcommand{\eq}{\begin{eqnarray}}
\newcommand{\en}{\end{eqnarray}}
\newcommand{\ra}{\rangle}

\begin{document}

\title{Strong decays of the hadronic molecule $\Omega^\ast (2012)$}

\author{Thomas Gutsche$^1$,
        Valery E. Lyubovitskij$^{1,2,3,4}$ 
\vspace*{1.2\baselineskip}}
\address{
$^1$ Institut f\"ur Theoretische Physik,
Universit\"at T\"ubingen, \\
Kepler Center for Astro and Particle Physics,
Auf der Morgenstelle 14, D-72076 T\"ubingen, Germany \\
$^2$ Departamento de F\'\i sica y Centro Cient\'\i fico
Tecnol\'ogico de Valpara\'\i so-CCTVal, Universidad T\'ecnica
Federico Santa Mar\'\i a, Casilla 110-V, Valpara\'\i so, Chile \\ 
$^3$ Department of Physics, Tomsk State University,
634050 Tomsk, Russia\\
$^4$ Tomsk Polytechnic University,
634050 Tomsk, Russia\\} 
\ead{
thomas.gutsche@uni-tuebingen.de,
valeri.lyubovitskij@uni-tuebingen.de 
}

\begin{abstract}

Strong two- and three-body decays of the new excited hyperon 
$\Omega^*(2012)$ are discussed in the hadronic molecular approach.   
The $\Omega^*(2012)$ state is considered to contain the mixed
$\Xi^* \bar K$ and $\Omega \eta$ hadronic components. 
In our calculations we use a phenomenological hadronic Lagrangian 
for description of interaction of the $\Omega^*(2012)$ state with constituents 
and of constituents to other hadrons occurring in the final state. 
Our results show that the decay widths of the two-body decay modes 
$\Omega^*(2012) \to \Xi \bar K$ lie in the few MeV region and are compatible 
with or dominate over the rates of the three-body modes 
$\Omega^*(2012) \to \Xi \pi \bar K$. The sum of two- and three-body 
decay widths is consistent with a width of the $\Omega^*(2012)$ 
originally measured by the Belle collaboration. 
A possible scenario for the suppression of the three-body decay rate
recently noticed by the Belle collaboration is due to the dominant admixture 
of the $\Omega \eta$ hadronic component in the $\Omega^*(2012)$ state. 

\end{abstract}

\date{\today}
\maketitle
\section{Introduction}
\label{introduction} 

Last year the Belle collaboration reported on a new excited isosinglet 
hyperon $\Omega^{*-}$ state decaying into $\Xi^0 K^-$ and $\Xi^- K_S^0$ 
pairs with a mass of $2012.4 \pm 0.7 \pm 0.6$ MeV and a width of 
$\Gamma = 6.4^{+2.5}_{-2.0} ({\rm stat}) 
\pm 1.6 ({\rm syst})$ MeV~\cite{Yelton:2018mag}. 
The spin-parity quantum numbers of the $\Omega^{*-}$ have been favored to be 
$J^P = \frac{3}{2}^{-}$ based on two arguments: 
(1) the observed mass value of the $\Omega^{*-}$ is close to 
the theoretical predictions for the $\frac{3}{2}^{-}$ states, 
(2) the rather narrow width of the $\Omega^{*-}$ 
decaying to a $\Xi \bar K$ pair via a $d$ wave. Recently, the Belle 
collaboration searched for the cascade three-body decay 
$\Omega(2012) \to \bar K \Xi(1530) \to \bar K \pi \Xi$~\cite{Jia:2019eav}. 
They did not observe any significant signal in this channel and derived 
upper limits for the ratios of the branchings relative to the 
two-body $\bar K \Xi$ decay modes. In particular, the most stringent upper 
limits read~\cite{Jia:2019eav}: 
\eq\label{R12} 
R_1 &=& \frac{B(\Omega(2012) \to \Xi(1530)^0(\to \Xi^-\pi^+)K^-)}
             {B(\Omega(2012) \to \Xi^-\bar K^0)} < 9.3 \%\,, 
\nonumber\\
& &\\
R_2 &=& \frac{B(\Omega(2012) \to \Xi(1530)^0(\to \Xi^-\pi^+)K^-)}
             {B(\Omega(2012) \to \Xi^0 K^-)} < 7.8 \%\,. \nonumber 
\en 
An excited $\Omega^*$ state with $J^P= \frac{3}{2}^{-}$ and a mass of 2020 MeV 
has been predicted in a quark model with a QCD based potential 
in Ref.~\cite{Chao:1980em}. Later, different types of 
quark models~\cite{QM_Omega}, large $N_c$ approaches~\cite{Carlson:2000zr}, 
algebraic string-like model~\cite{Bijker:2000gq}, 
Skyrme model~\cite{Oh:2007cr}, and lattice QCD~\cite{Engel:2013ig} 
reported on an estimate for the $\Omega^*$ mass in 
the region from 1953 to 2120 MeV. Inclusion of sizeable five-quark 
Fock components and their mixing with three-quark components in 
the constituent quark models, considered in Refs.~\cite{Yuan:2012zs}, 
lead to a reduction of the $\Omega^*$ mass by about 200 MeV. 
The dynamical generation of the $\Omega^*$ state using two ($\Omega \eta$ 
and $\Xi^* \bar K$) coupled channels in the chiral unitary approach 
has been proposed 
and developed in Refs.~\cite{Kolomeitsev:2003kt}-\cite{Si-Qi:2016gmh}, 
where the mass of the $\Omega^*$ has a strong dependence on the choice of 
model parameters leading to 2141 MeV in Ref.~\cite{Sarkar:2004jh} and 
1800 MeV in Ref.~\cite{Si-Qi:2016gmh}. 

The understanding of the structure and decays of the $\Omega^*(2012)$ 
state has been of increased interest since the discovery by the Belle 
collaboration. In Refs.~\cite{Xiao:2018pwe}-\cite{Lin:2019tex} possible 
interpretations of 
the $\Omega^*(2012)$ hyperon have been critically discussed. 
In Ref.~\cite{Xiao:2018pwe} the two-body decays $\Omega^* \to \Xi^0 K^-$ and 
$\Omega^* \to \Xi^- \bar K^0$ have been analyzed in the chiral quark model. 
It was argued that the obtained numerical results are in agreement with 
spin-parity $\frac{3}{2}^-$ of the $\Omega^*$ state, 
while alternative assignments as $\frac{1}{2}^-$ and 
$\frac{3}{2}^+$ cannot be completely excluded. In Ref.~\cite{Aliev:2018syi} 
the possible structure and resulting strong decays of the $\Omega^*$
were studied using QCD 
sum rules. From the analysis of the mass and the strong decay properties (coupling  
constants and widths) it was concluded that the $\Omega^*$ hyperon is the 
$1P$ orbital excitation of the ground state $\Omega(1670)$ baryon with 
$J^P = \frac{3}{2}^{-}$. The same conclusion about the nature of the $\Omega^*$ 
state has been made from an analysis of its two-body decays in the framework of 
the $^3P_0$ model. In Ref.~\cite{Polyakov:2018mow} the $\Omega^*$ state has 
been considered on the basis of the $SU(3)$ flavor picture. 
It was found that if the $\Omega^*$ state is the $\bar K \Xi(1530)$ molecular 
state formed in the isospin zero channel, then its main decay mode is the 
tree-body process $\Omega^* \to \Xi \bar K \pi$. 
In Ref.~\cite{Valderrama:2018bmv} a hadronic molecular scenario for 
the $\Omega(2012)$ has been tested in an effective field-theoretical approach. 
It was found that the partial width of the three-body 
decay $\Omega^* \to \Xi \bar K \pi$ is in the 2-3 MeV interval, 
while the partial width of the two-body decay $\Omega^* \to \Xi \bar K$ 
is in the 1-11 MeV range. Here and also 
in Refs.~\cite{Lin:2018nqd,Huang:2018wth,Pavao:2018xub}  
the dominance or sizable contribution of the tree-body 
decays $\Omega^* \to \Xi \bar K \pi$ has been based on the description of 
these processes by a tree-level diagram, while the two-body processes 
$\Omega^* \to \Xi \bar K$ have been described by a loop-diagram generated 
by the $\Omega^*$ constituents. 
In Ref.~\cite{Liu:2019wdr} the $\Omega$ baryon spectrum up to  the 
$N=2$ shell has been calculated based on 
a nonrelativistic constituent quark potential model. 
In Ref.~\cite{Lin:2019tex} the analysis of the $\Omega(2012)$ 
strong decays has been performed using a hadronic molecular model 
and different spin-parity assignments for these state. 

The formalism of quantum field theory for treatment of 
bound states based on the Weinberg-Salam (WS) compositeness condition. 
It was formulated in Refs.~\cite{Salam:1962ap}-\cite{Efimov:1993ei} 
and widely applied to the study of hadronic molecules (HM) 
in Refs.~\cite{Salam:1962ap} 
and~\cite{Faessler:2007gv}-\cite{Dong:2008gb}.  
Application of the WS condition for description of composite 
structure of conventional hadrons~\cite{Salam:1962ap}-\cite{Efimov:1993ei} and 
exotic states~\cite{Salam:1962ap},\cite{Faessler:2007gv}-\cite{Dong:2008gb} 
is based on the following steps:  
(1) First, we derive a phenomenological Lagrangian, which is manifestly 
Lorenz covariant and gauge invariant and describes the interaction of the bound 
state and its constituents; 
(2) Second, the coupling constant of the HM with 
the constituents is fixed by solving by the WS condition 
$Z_{\rm HM} = 1 - \Pi_{\rm HM}' = 0$~\cite{Salam:1962ap}-\cite{Dong:2008gb}, 
where $\Pi_{\rm HM}'$ is the derivative of the HM mass operator induced by 
interaction Lagrangian of the HM with its constituents. 
The WS condition is understood as the probability 
to find the HM as the bare state is always equal to zero; 
(3) Third, we construct the $S$-matrix operator, which consistently generates matrix 
elements of physical processes involving HM. 
By construction we are able
to consider hadronic molecules as an admixture of two or more 
hadronic constituent channels (Fock states). This procedure corresponds 
to the consideration of 
two- or multiple channel couplings in potential approaches. Of course, 
the use of a mixture of hadronic constituents leads to 
additional parameters, which reduce the  predictive power of the approach. 
Therefore, one can try to reduce parameters by using additional physical 
arguments, like, e.g., the hadronic compoments whose sum of the masses 
is far away from the mass of the hadronic molecule have strongly reduced 
contributions. We therefore restrict to the leading components. 

The main our goal is to present a self-consistent piciture of strong two- and 
three-body decays of the $\Omega^*(2012)$ state in the hadronic molecular 
approach formulated and developed in~\cite{Faessler:2007gv}-\cite{Dong:2008gb}.   
In the recent paper by the Belle collaboration~\cite{Jia:2019eav} it is 
claimed that the strong three-body decay modes of the $\Omega(2012)$ 
state are suppressed in comparison with the two-body one, which
raises doubts on a molecular interpretation of this state.
We find that due to a possible mixture of the hadronic components
$\Xi^* \bar K$ and $\Omega \eta$ in the structure of $\Omega$ 
the three-body decays $\Omega(2012) \to \Xi \pi \bar K$ are suppressed 
when the $\Omega \eta$ hadronic component dominates. 
Note that applications of our approach to hadrons
composed of two and more hadronic configurations have been 
performed in Refs.~\cite{Dong:2008gb,Branz:4430,Dong:2940} 
with respect to the $X(3872)$, $Z(4430)$, and $\Lambda(2940)$ states. 

\section{Hadronic molecular structure of the $\Omega(2012)$}  
\label{framework} 

Following the conjecture of the Belle collaboration~\cite{Yelton:2018mag} and 
assignments of most of the theoretical approaches we accept that 
$\Omega^*(2012)$ has $J^{P} = \frac{3}{2}^{-}$ spin-pariry. 
We consider the $\Omega^*(2012)$ as a weakly bound hadronic molecule, which
involves a superposition of two hadronic components --- 
$(\Omega[1670] \eta)$ and $(\Xi^*[1530] \bar K$): 
\eq 
|\Omega^*(2012) \ra = \cos\theta \, 
\frac{ | \Xi^{* 0} K^- \ra \, + \, | \Xi^{* -} \bar K^0 \ra }{\sqrt{2}} 
- \sin\theta \, |\Omega^- \eta \ra \,. 
\en 
A mixing angle $\theta $ is introduced between the two components, which
later on will play an 
important role in explaining a possible suppression of the three-body decays 
$\Omega(2012) \to \Xi \pi \bar K$ recently observed by the Belle 
collaboration~\cite{Jia:2019eav}. Such angle can be constrained from data 
on the $\Omega^*(2012)$ decays. In particular, in this paper we show 
that a possible scenario for the dominance of the two-body decay rates over 
three-body ones recently noticed by the Belle collaboration could occur due to 
the dominant admixture 
of the $\Omega \eta$ hadronic component in the $\Omega^*(2012)$ state.  

For convenience, 
in Table I we present the quantum numbers (isospin $I$, spin-parity $J^P$) 
and values of masses (current central values from the Particle Data 
Group~\cite{PDG}) of the hadrons, which will be used in our calculations.
Note that our approach is able to describe the molecular states as mixing 
of different Fock commponents and before it was successfully applied 
to the problem of the $X(3872)$ state~\cite{Dong:2008gb}. 

\begin{table}[hb]
\begin{center}
\caption{Quantum numbers and masses of relevant hadrons}

\vspace*{.15cm}

\def\arraystretch{1.4}
\begin{tabular}{c|c|c|c}
\hline
\ \ Hadron \ \ & \ \ $I$ \ \ & \ \ $J^P$ \ \ & \ \ Mass (MeV) \ \ \\
\hline
$\pi^\pm$      & $1$             & $0^-$ & 139.57061 \\
\hline
$\pi^0$        & $1$             & $0^-$ & 134.977   \\
\hline
$\bar K^0$     & $\frac{1}{2}$   & $0^-$ & 493.677   \\
\hline
$K^-$          & $\frac{1}{2}$   & $0^-$ & 497.611   \\
\hline
$\eta$         & $0$             & $0^-$ & 547.862   \\
\hline
$\bar K^{* -}$ & $\frac{1}{2}$   & $1^-$ & 891.76    \\
\hline
$K^{* 0}$      & $\frac{1}{2}$   & $1^-$ & 895.55    \\
\hline
$\Xi^{0}$      & $\frac{1}{2}$   & $\frac{1}{2}^+$ & 1314.86 \\
\hline
$\Xi^{-}$      & $\frac{1}{2}$   & $\frac{1}{2}^+$ & 1321.71 \\
\hline
$\Xi^{* 0}$    & $\frac{1}{2}$   & $\frac{3}{2}^+$ & 1531.8 \\
\hline
$\Xi^{* -}$    & $\frac{1}{2}$   & $\frac{3}{2}^+$ & 1535.0 \\
\hline
$\Omega^{-}$   & $0$             & $\frac{3}{2}^+$ & 1672.45 \\
\hline
$\Omega^{* -}$ & $0$             & $\frac{3}{2}^-$ & 2012.4 \\
\hline
\end{tabular}
\label{tab:hadrons}
\end{center}
\end{table}

The interaction of the $\Omega^*(2012)$ state and their constituents 
is described by the phenomenological Lagrangian: 

\eq\label{LOmega_int}
{\cal L}_{\Omega^*}(x) &=& g_{\Omega^*} \, \bar\Omega^*_\mu(x) \,
\int d^4y \, \Phi(y^2) \,
\biggl[ \frac{\cos\theta}{\sqrt{2}} \, \Xi^{*}_i(x+\omega_{K} y) 
\bar K_i(x-\omega_{\Xi^*}y) \nonumber\\
&-& \sin\theta \, \Omega(x+\omega_\eta y) 
\eta(x-\omega_\Omega y) \biggr] \, + \, {\rm H.c.}
\en
Here $\Xi_i^{*} = (\Xi^{* 0}, \Xi^{* -})$ and  
$\bar K_i = (K^-, \bar K^0)$ are the doublets of 
$\Xi^{*}$ hyperons and $\bar K$ mesons, 
$\Phi(y^2)$ is the vertex form factor modeling the
distribution of $(\Xi^{*} \bar K)$ and $(\Omega \eta)$ constituents
in the $\Omega^*$ state, $\omega_{\Xi^*} = M_{\Xi^*}/(M_{\Xi^*}+M_K)$,  
$\omega_{K} = M_K/(M_{\Xi^*}+M_K)$, 
$\omega_{\Omega} = M_{\Omega}/(M_{\Omega}+M_\eta)$, and 
$\omega_{\eta} = M_{\eta}/(M_{\Omega}+M_\eta)$ are the fractions 
of the constituent masses obeying the conditions 
$\omega_{\Xi^*} + \omega_{K} = 1$ and 
$\omega_{\Omega} + \omega_\eta = 1$. 
For the Fourier transform of the $\Phi(y^2)$
we use 
$\tilde\Phi(p^2_E) \doteq \exp(-p^2_E /\Lambda^2)$, 
where $p_E$ is the relative Jacobi momentum in Euclidean space 
and $\Lambda$ is the dimensional parameter, which has a value of about 1 GeV. 
The cutoff parameter $\Lambda$ characterizes distribution of the 
bound state $\Omega^*(2012)$. Hence, $\Lambda$ has a universal value 
independent of the parameter decay process. 

Note that the $|\Omega^*(2012)>$ state could also contain an admixture 
of a three $s$-quark component. In such a case the Fock state and 
the corresponding phenomenological Lagrangian describing the coupling 
of the $\Omega^*(2012)$ to the  constituents are modified as 
\eq 
\hspace*{-2cm}
|\Omega^*(2012) \ra = \cos\phi \, \biggl[ \cos\theta \, 
\frac{ | \Xi^{* 0} K^- \ra \, + \, | \Xi^{* -} \bar K^0 \ra }{\sqrt{2}} 
- \sin\theta \, |\Omega^- \eta \ra \biggr] + \sin\phi \, |sss\ra  
\en 
and 
\eq 
\hspace*{-2cm}
{\cal L}_{\Omega^*}(x) &=& g_{\Omega^*} \, \bar\Omega^*_\mu(x) \,
\int d^4y \, \biggl\{ \Phi(y^2) \, \cos\phi 
\biggl[ \frac{\cos\theta}{\sqrt{2}} \, \Xi^{*}_i(x+\omega_{K} y) 
\bar K_i(x-\omega_{\Xi^*}y) \nonumber\\
\hspace*{-2cm}
&-& \sin\theta \, \Omega(x+\omega_\eta y) 
\eta(x-\omega_\Omega y) \biggr] 
\, + \, \int d^4z \, \Phi(y^2+z^2) \, \sin\phi \, \epsilon^{abc}
\, s^a(x + \xi_+) \nonumber\\
\hspace*{-2cm}
&\times& s^b(x+ \xi_-) \, C \, \gamma^\mu \, s^c(x- \xi_+ - \xi_-) 
\biggr\}
\, + \, {\rm H.c.} \,, \qquad 
\xi_\pm = \frac{y}{3 \sqrt{2}} \pm \frac{z}{\sqrt{6}}\,.    
\en 
Now $\phi$ is an additional parameter, which defines the mixing 
of hadronic and quark components in the $\Omega^*(2012)$. 
In this paper we proceed with a hadronic molecular scenario.  
The possible mixing of the hadronic molecular and quark 
components will be considered in an extended work. 
One should stress that our approach 
was already successfully applied to the mixing 
of hadronic and quark components in exotic bound states. 
In particular, in a series of papers~\cite{Dong:2008gb} 
the exotic state $X(3872)$ was considered in the mixing 
of a charm-anticharm component and a set of hadronic
components $(DD^*)$, $(J\psi \rho)$, and $(J\psi \omega)$, 
while the decay properties were successfully described. 

\begin{figure}[hb]
\begin{center}
\epsfig{figure=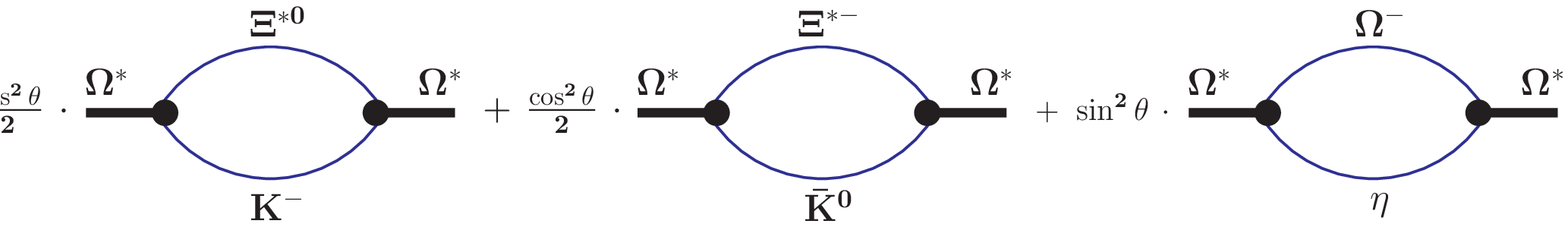,scale=.7}
\caption{Diagrams representing the mass operator of the $\Omega^*(2012)$ 
state.} 
\label{fig:Mass_M}
\end{center}
\end{figure}

The coupling $g_{\Omega^*}$ is fixed determined using the compositeness
condition~\cite{Salam:1962ap}-\cite{Dong:2008gb}  
\eq\label{ZOmega}
Z_{\Omega^*} = 1 - \Sigma_{\Omega^*}^{T \prime}(M_{\Omega^*}) \equiv 0 \,,
\en
where $\Sigma_{\Omega^*}^{T \prime}$ is the derivative of the transversal 
contribution to 
The $\Omega^*(2012)$ mass operator: 
\eq
\Sigma_{\Omega^*}^{\mu\nu}(p) =
  g_{\perp}^{\mu\nu} \Sigma_{\Omega^*}^T(p) 
+ \frac{p^\mu p^\nu}{p^2} \Sigma_{\Omega^*}^L(p),
\en
with $g^{\mu\nu}_{\perp}=g^{\mu\nu}-p^\mu p^\nu /p^2$. Note, the longitudinal part 
of the mass operator vanishes due to the Lorenz transversality condition  
$p^\mu \, \epsilon_\mu(p) = 0$, where $\epsilon_\mu(p)$ is the polarization vector 
of the $\Omega(2012)$ state.  
The corresponding Feynman diagrams contributing to the mass operator 
$\Sigma_{\Omega^*}$, which are generated by the loops of the 
$(\Xi^{* 0} K^-)$, $(\Xi^{* -} \bar K^0)$, and $(\Omega^- \eta)$ 
constituents, are displayed in Fig.~\ref{fig:Mass_M}. 
In addition we have a free parameter $\theta$, which is the
mixing angle between the $\Xi^* \bar K$ and the $\Omega \eta$ hadronic
molecular components of the $\Omega^*$ state. 
One should stress that the mass operator of the $\Omega^*(2012)$ state 
is represented by only two loop diagrams in terms of two types 
of the constituents $\Xi^* \bar K$ and $\Omega \eta$. 
It is the diagonalized representation of the compositeness condition, 
which is fully equivalent to
the formalism used in Refs.\cite{Si-Qi:2016gmh,Huang:2018wth,Pavao:2018xub} 
where the bound state is generated by scattering processes of its constituents, which includes 
both diagonal $\Xi^* \bar K \to \Xi^* \bar K$ and $\Omega \eta \to \Omega \eta$ 
and non-diagonal $\bar K\Xi^* \to \eta\Omega$ transitions.

The expression for the mass operator $\Sigma_{\Omega^*}$ is given by 
\eq
& &\Sigma_{\Omega^*}^{\mu\nu}(p) = \frac{\cos^2\theta}{2} 
\, \biggl(\Sigma_{\Omega^*}^{\mu\nu, \Xi^0K^-}(p) 
      + \Sigma_{\Omega^*}^{\mu\nu, \Xi^-\bar K^0}(p)\biggr)
\, + \, \sin^2\theta \, \Sigma_{\Omega^*}^{\mu\nu, \Omega\eta}(p)\,, 
\label{SOmegastar}\nonumber\\
& &\Sigma_{\Omega^*}^{\mu\nu, H_1H_2}(p) = 
g_{\Omega^*}^2 \,
\int \frac{d^4k}{(2\pi)^4 i} \tilde\Phi^2(-(k+p\omega_2)^2) \,
S_{H_1}^{\mu\nu}(k+p) \, S_{H_2}(k) \,, 
\label{SBoson}
\en
where 
\eq 
S_{H_1}^{\mu\nu}(k) &=& \frac{1}{M_{H_1} - \not\! k} \, 
\biggl[ - g^{\mu\nu} + \frac{\gamma^\mu \gamma^\nu}{3} 
+ \frac{2 k^\mu k^\nu}{3 M_{H_1}^2} 
+ \frac{\gamma^\mu k^\nu - \gamma^\nu k^\mu}{3 M_{H_1}} \biggr]\,, \\
S_{H_2}(k) &=& \frac{1}{M_{H_2}^2 - k^2} 
\en 
are the propagators of 
baryon $H_1$ with spin $\frac{3}{2}$ and of meson $H_2$ 
with spin $0$. Here $H_1 = \Xi^{* -}, \Xi^{* 0}, \Omega^-$ 
and $H_2 = K^-, \bar K^0, \eta$. 
 
The coupling constant $g_{\Omega^*}$ deduced from the compositeness 
condition~(\ref{ZOmega}) reads 
\eq\label{coupling_Omega}
\frac{g_{\Omega^*}^2}{16\pi^2} &=& \frac{1}{I_{\Omega^*}}\,,
\en
where $I_{\Omega^*}$ is the structure integral
\eq
& &I_{\Omega^*} = \frac{\cos^2\theta}{2} \, 
\biggl(I_{\Omega^*}^{\Xi^{* 0}K^-} + I_{\Omega^*}^{\Xi^{* -}\bar K^0}\biggr)
\, + \, \sin^2\theta \ I_{\Omega^*}^{\Omega \eta} \,, \nonumber\\
& &I_{\Omega^*}^{H_1H_2} =
\int\limits_0^\infty \, \frac{d\alpha_1 d\alpha_2}{\Delta^3} \, 
R_{\Omega^*} \, e^{-w_{\Omega^*}} \,, \quad \Delta = 2  + \alpha_1 + \alpha_2  
\en
and 
\eq 
\hspace*{-2cm}
& &R_{\Omega^*} = \Big(1 + \frac{1}{3 \mu_{H_1^2} \Delta} \Big) 
\ \biggl[ \alpha_2 + 2 \omega_1 + 2 \mu_{\Omega^*} \Big(\mu_{H_1} 
+ \mu_{\Omega^*} \frac{\alpha_2 + 2 \omega_1}{\Delta} \Big) 
\, z(\alpha_1,\alpha_2) 
\biggr]
\,, \nonumber\\
\hspace*{-2cm}
& &w_{\Omega^*} = \frac{(\mu_{H_1}\alpha_1 - \mu_{H_2}\alpha_2)^2}{\Delta} 
+ \frac{(\mu_{H_1}+\mu_{H_2})^2 - \mu_{\Omega^*}^2}{\Delta} \, 
z(\alpha_1,\alpha_2) \,, \nonumber\\
\hspace*{-2cm}
& &\mu_i = \frac{M_i}{\Lambda}\,, \qquad z(\alpha_1,\alpha_2) 
= \alpha_1\alpha_2 + 2 \alpha_1\omega_1^2 + 2 \alpha_2\omega_2^2 \,. 
\en

\begin{figure}
\begin{center}
\epsfig{figure=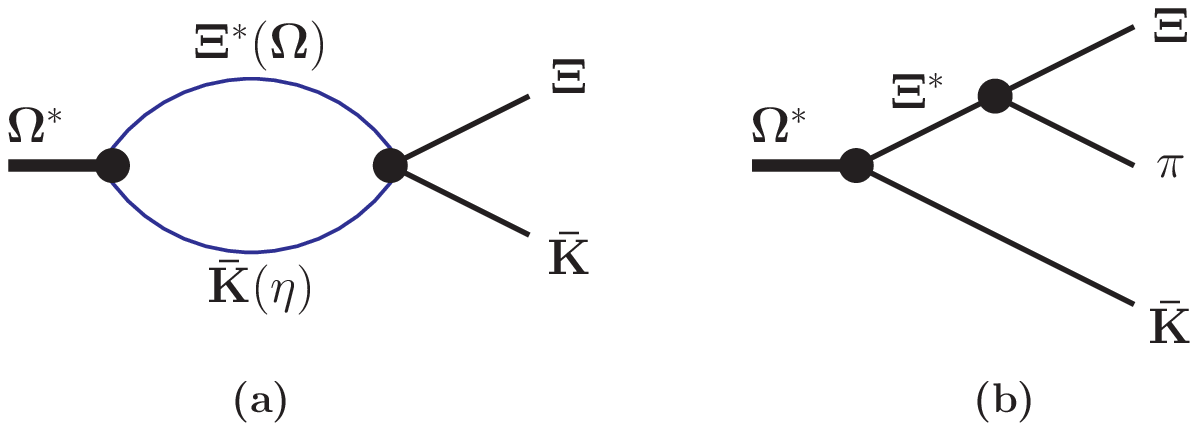,scale=.8}
\caption{Leading diagrams contributing to the strong decays of the $\Omega^*$ 
state: (a) two-body decay $\Omega^* \to \Xi \bar K$, (b) three-body decay 
$\Omega^* \to \Xi \pi \bar K$.} 
\label{fig:Decays_M}
\end{center}
\end{figure}

The leading diagrams contributing to the strong decays of the $\Omega^*$ state are 
shown in Fig.~\ref{fig:Decays_M}: (a) two-body decay $\Omega^* \to \Xi \bar K$, 
(b) three-body decay $\Omega^* \to \Xi \pi \bar K$. Note that 
in agreement with Refs.~\cite{Polyakov:2018mow,Valderrama:2018bmv,%
Lin:2018nqd,Huang:2018wth,Pavao:2018xub} the two-body decay 
$\Omega^* \to \Xi \bar K$ proceeds via the hadronic loops $(\Xi^* \bar K)$ and 
$(\Omega \eta)$ involving the constituents of the $\Omega^*$ state. 
The three-body decay 
$\Omega^* \to \Xi \pi \bar K$ is described by the two-cascade tree-level diagram, 
where the $\Omega^*$ first couples to the constituents $\Xi^*$ and 
$\bar K$ and then $\Xi^*$ decays via the dominant mode into $\Xi$ and $\pi$. 
For calculation of the diagrams in Fig.~\ref{fig:Decays_M} we built 
the Lagrangian including the interaction of the $\Omega^*$ with its hadronic 
constituents $\Xi^* \bar K$ and $\Omega \eta$, and two additional terms 
describing the couplings $\Xi^* \Xi K K \, (\Xi \Omega \eta K)$ and 
$\Xi^* \Xi \pi$.

The $\Xi^* \Xi \pi$ interaction is described by the phenomenological Lagrangian 
\eq 
{\cal L}_{\Xi^*\Xi\pi} = \frac{g_{\Xi^*\Xi\pi}}{M_{\Xi^*}} \, 
\bar\Xi^{*}_\mu \partial^\mu\vec{\pi} \vec{\tau} \, \Xi \, + \, {\rm H.c.}\,, 
\en 
where the dimensionless  coupling $g_{\Xi^*\Xi\pi}$ is fixed from data 
on the $\Xi^* \to \Xi \pi$ decay width. In particular, 
the corresponding two-body decay width reads 
\eq 
\Gamma(\Xi^* \to \Xi \pi) = \frac{g_{\Xi^*\Xi\pi}^2}{64 M_{\Xi^*}^7} \, 
\lambda^{3/2}(M_{\Xi^*}^2,M_\Xi^2,M_\pi^2) \, 
\Big[ (M_{\Xi^*}+M_\Xi)^2 - M_\pi^2 \Big] \,. 
\en 
Using the measured central values of 
\eq 
& &\Gamma_{\rm total}(\Xi^{* 0}) = 2 \Gamma(\Xi^{* 0} \to \Xi^- \pi^+) 
+ \Gamma(\Xi^{*0} \to \Xi^0 \pi^0) = 9.14 \ {\rm MeV}\,, \nonumber\\ 
& &\Gamma_{\rm total}(\Xi^{* -}) = 2 \Gamma(\Xi^{* -} \to \Xi^0 \pi^-) 
+ \Gamma(\Xi^{*-} \to \Xi^- \pi^0) = 9.9 \ {\rm MeV} 
\en 
one gets $g_{\Xi^{*}\Xi\pi} = 6.79$ from the $\Xi^{* 0}$ set and 
$g_{\Xi^{*}\Xi\pi} = 6.71$ from $\Xi^{* -}$. In the following we will 
use the averaged value $g_{\Xi^{*}\Xi\pi} = (6.79+6.71)/2 = 6.75$.  

The couplings $\Xi^* \Xi K K$ and $\Xi \Omega \eta K$ are generated by 
a phenomenological Lagran\-gian:  
\eq\label{GammaDB}
{\cal L}_{\Gamma DB} = g_{\Gamma DB} \, 
\bar B^{mk} \, \gamma^5 \, \Gamma^\mu_{lj} \, D^{ijk}_\mu 
\, \epsilon^{ilm} \, + \, {\rm H.c.}\,,  
\en
where 
$B^{mk}$ and $D^{ijk}$ are the octet and decuplet baryon fields, 
$\epsilon^{ilm}$ is the rang-3 Levi-Civita tensor.   
$\Gamma_\mu$ is the chiral connection which in absence of 
external vector and axial fields is defined as 
\eq 
\Gamma_\mu = \frac{1}{2} [u^+, \partial_\mu u] = 
\frac{1}{4 F^2} [\hat\Phi, \partial_\mu \hat{\Phi}] + 
{\cal O}(\hat\Phi^4)\,, 
\en 
where $\hat\Phi = \sum\limits_{i=1}^8 \phi_i \lambda_i$ 
is the octet matrix of pseudoscalar mesons. 

The dimensionless coupling $g_{\Gamma DB}$ can be fixed using the following 
arguments. We constraint $g_{\Gamma DB}$ based on the use of SU(2) chiral Lagrangians 
involving pseudoscalar, spin-$\frac{1}{2}$ and spin-$\frac{3}{2}$ fields and 
apply the extension to the SU(3) case~\cite{Pascalutsa:2006up,Ledwig:2014rfa,Blin:2015era}. 
In the SU(2) case, the $g_{\Gamma N\Delta}$ coupling is obtained by comparing 
it to three similar couplings describing the coupling of the two 
nucleon and nucleon-$\Delta$ pair to the pion ($g_{\pi NN}$, 
$g_{\pi N\Delta} = f_{\pi N\Delta} \, M_N/M_\pi$) and 
of two nucleons with the chiral connection ($g_{\Gamma NN} = 1$). 
On phenomenological grounds  
we postulate that all four couplings are related as: 
\eq\label{Couplings_Ansatz} 
\frac{g_{\Gamma N\Delta}}{g_{\Gamma NN}} = \frac{g_{\pi N\Delta}}{g_{\pi NN}} \,.
\en   
From the ansatz~(\ref{Couplings_Ansatz}) one gets 
\eq 
g_{\Gamma N\Delta} = g_{\Gamma NN} \, \frac{g_{\pi N\Delta}}{g_{\pi NN}} 
= \frac{f_{\pi N \Delta}}{g_{\pi N N}} \, 
\frac{M_N}{M_\pi} \,. 
\en 
Next using following fundamental constants of hadron physics: 
$\pi NN$ coupling $g_{\pi N N} \simeq 13.4$ and the $\pi N \Delta$ coupling 
$f_{\pi N \Delta} \simeq 2$ we obtain $g_{\Gamma N\Delta} \simeq 1$. 
Extending our considerations from SU(2) to SU(3) we finally arrive at the coupling 
$g_{\Gamma BD} \simeq 1$ defining the Lagrangian in Eq.~(\ref{GammaDB}), 
whose value will be used in our numerical analysis. 

\begin{figure}
\begin{center}
\epsfig{figure=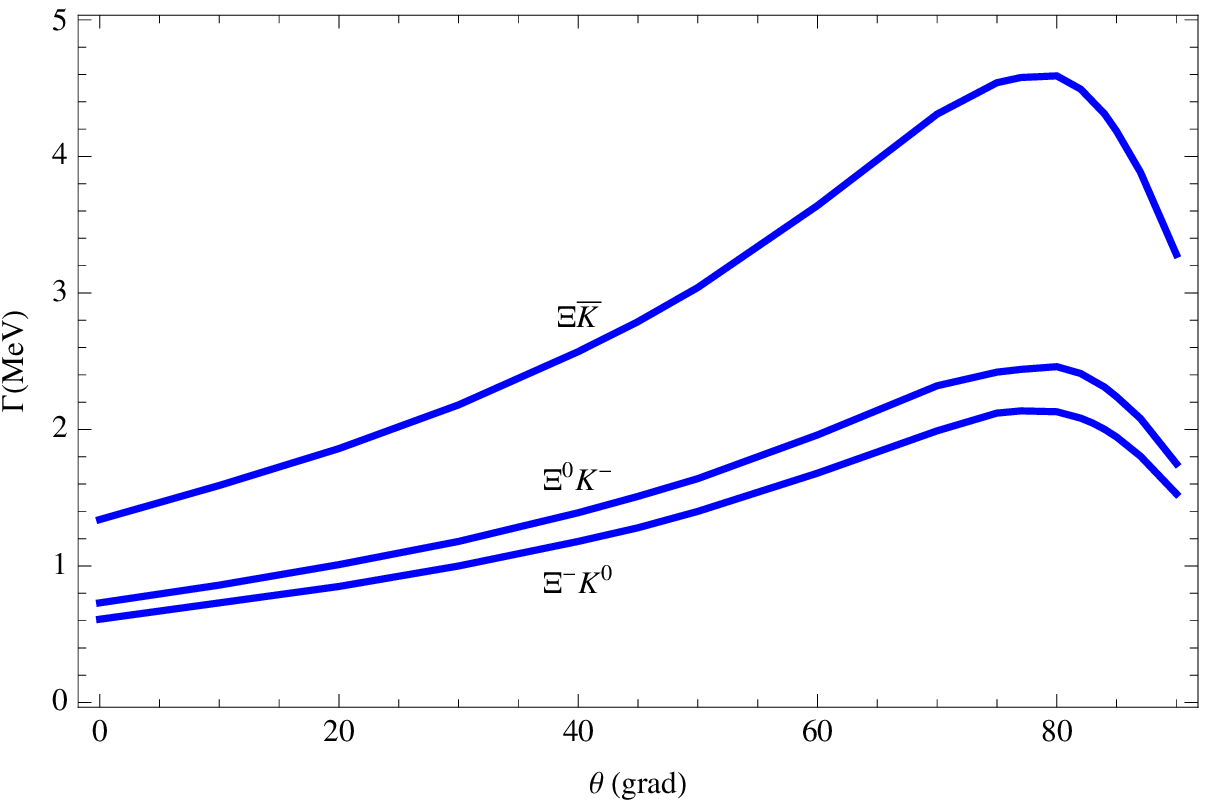,scale=.68}
\epsfig{figure=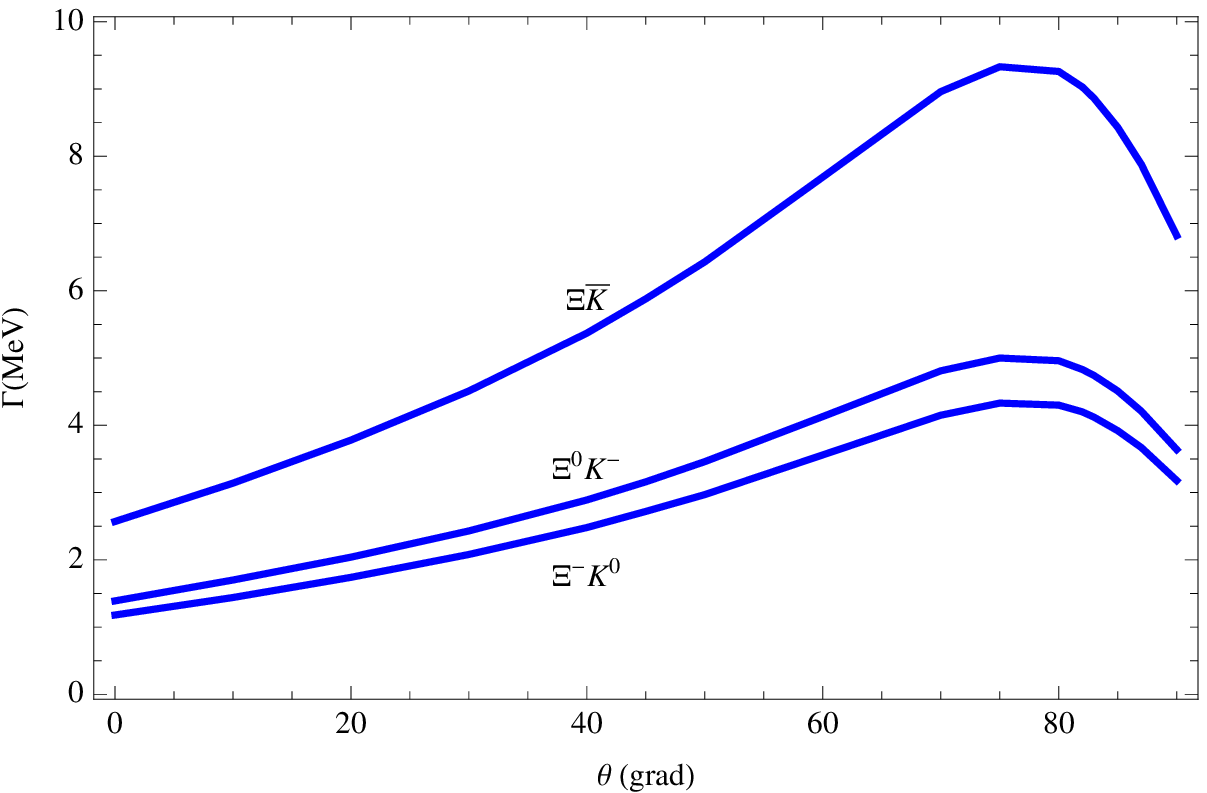,scale=.68}
\caption{Two-body decay widths for 
$\Omega^* \to \Xi \bar K$  
at $\Lambda = 1$ GeV (left panel) 
and $\Lambda = 1.5$ GeV (right panel).} 
\label{fig:res1}
\end{center}
\end{figure}

\begin{figure}
\begin{center}
\epsfig{figure=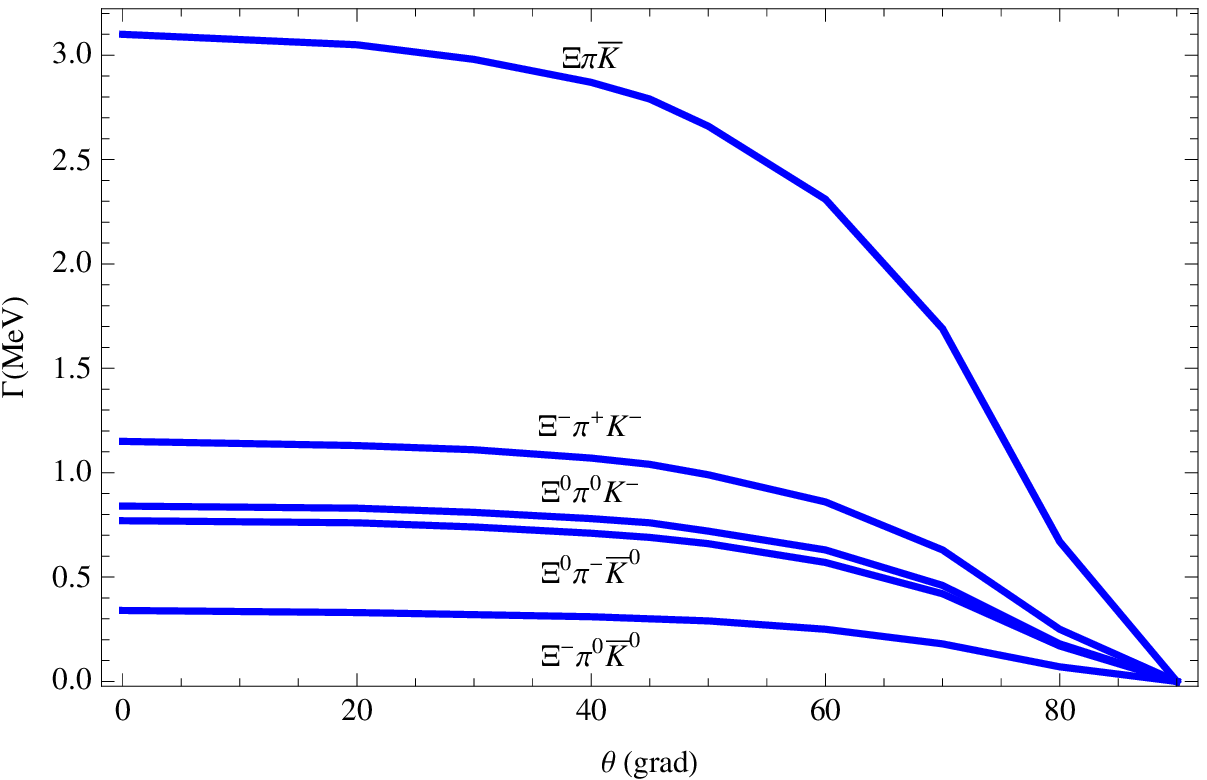,scale=.68}
\epsfig{figure=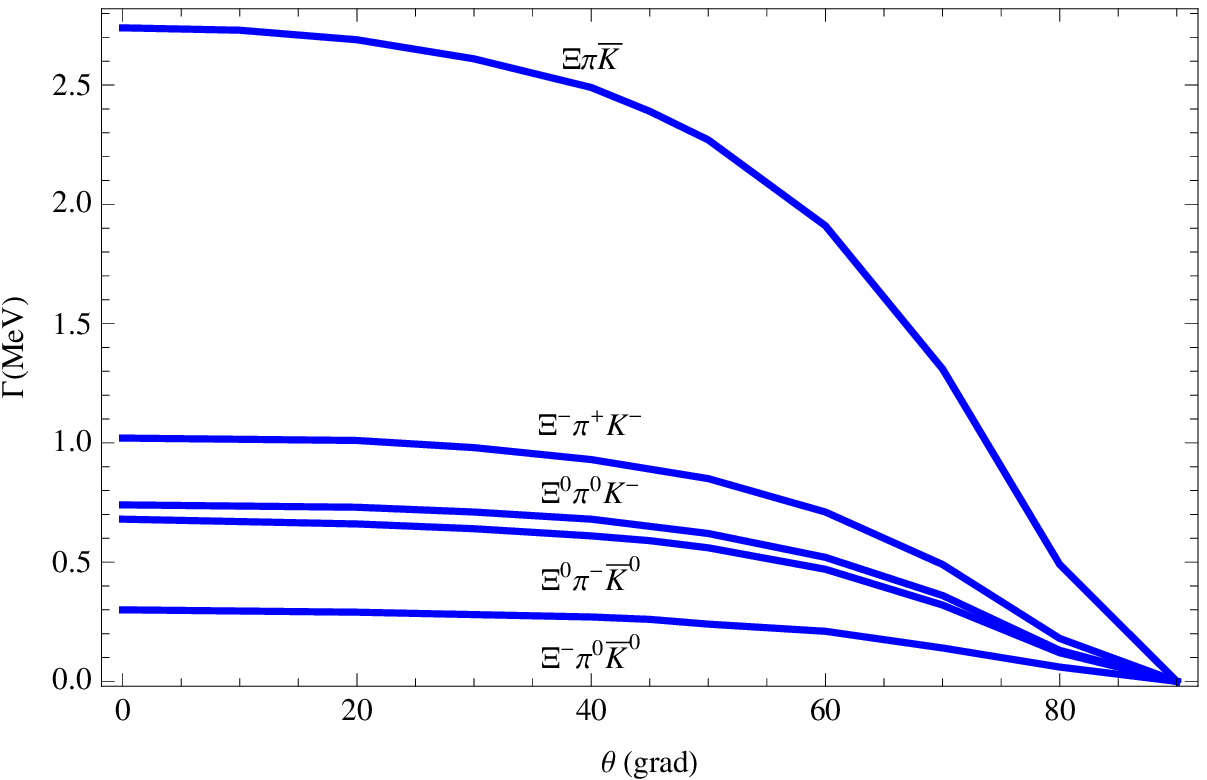,scale=.68}
\caption{Three-body decay widths for 
$\Omega^* \to \Xi \pi \bar K$ 
at $\Lambda = 1$ GeV (left panel) 
and $\Lambda = 1.5$ GeV (right panel).} 
\label{fig:res2}
\end{center}
\end{figure}

\begin{figure}
\begin{center}
\epsfig{figure=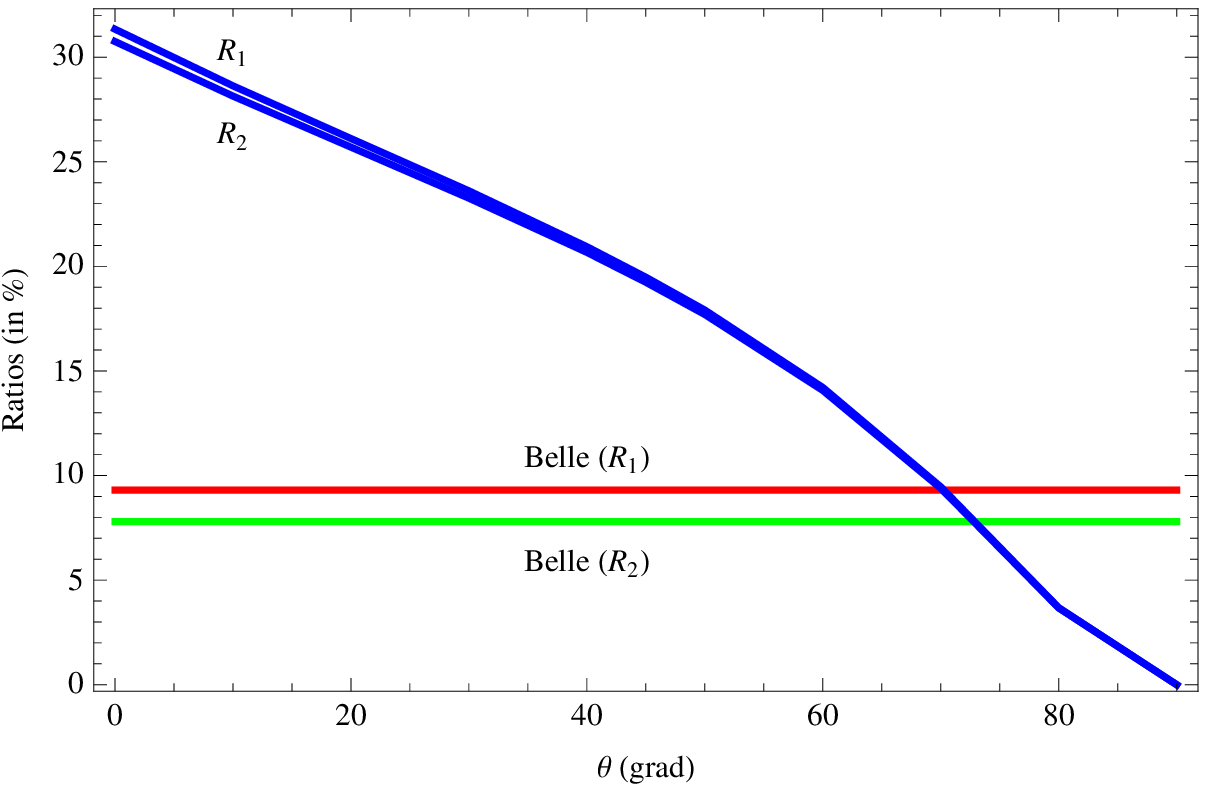,scale=.68}
\epsfig{figure=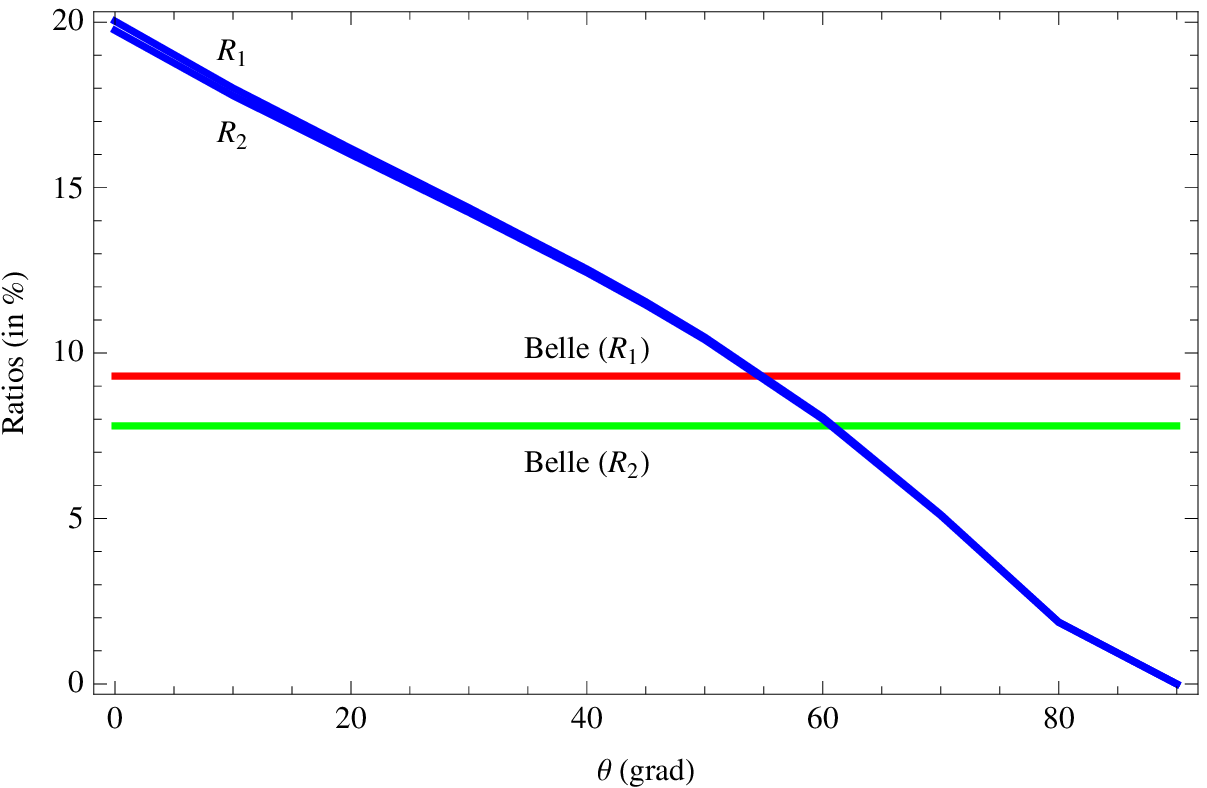,scale=.68}
\caption{Ratios $R_1$ and $R_2$ in comparison with the 
Belle upper limits of Eq.~(\ref{R12}) at at $\Lambda = 1$ GeV (left panel) 
and $\Lambda = 1.5$ GeV (right panel).} 
\label{fig:res3}
\end{center}
\end{figure}

The two- and three-body decay widths of the $\Omega^*$ state are calculated 
according to the standard expressions: 
\eq 
\hspace*{-2cm}
\Gamma(\Omega^* \to \Xi \bar K) &=& \frac{g_{\Omega^* \Xi K}^2}{192 \pi 
M_{\Omega^*}^3} 
\, \lambda^{3/2}(M_{\Omega^*}^2,M_\Xi^2,M_K^2) \, 
\Big[ (M_{\Omega^*} - M_\Xi)^2 - M_K^2 \Big]\,, \label{twobody_dec}\\ 
\hspace*{-2cm}
\Gamma(\Omega^* \to \Xi \pi \bar K) &=& \frac{1}{1024 \pi^3 M_{\Omega^*}^3} \, 
\int\limits_{(M_\Xi+M_\pi)^2}^{(M_{\Omega^*} - M_K)^2} ds_2 \, 
\int\limits_{s_1^-}^{s_1^+} ds_1 \, \sum\limits_{\rm pol} 
\Big|M_{\rm inv}\Big|^2 \,. 
\en 
$M_{\rm inv}$ is the invariant matrix element for the 
$\Omega^* \to \Xi \pi \bar K$ transition given by 
\eq\label{Minv3} 
\hspace*{-2cm}
M_{\rm inv} = 
g_{\Omega^*} \, \Phi\biggl(-[p_1 w_{\Xi^*} -  q w_K]^2\biggr)   
\, \frac{\cos\theta}{\sqrt{2}} \, \frac{g_{\Xi^*\Xi\pi}}{M_{\Xi^*}} \, c_\pi \, 
\bar u_\Xi(q) \, S_{\Xi^*; \mu\nu}(q) \, u_{\Omega^*}^\nu(p) \, p_3^\mu  
\en 
and the $\Omega^* \Xi K$ coupling $g_{\Omega^* \Xi K}$ 
is evaluated from the diagram in Fig.~\ref{fig:Decays_M}a. 
Here we introduced the following notations. 
$s_1$ and $s_2$ are the invariant Mandelstam variables 
$s_1 = (p_1 + p_2)^2 = (p - p_3)^2$ and 
$s_2 = (p_2 + p_3)^2 = (p - p_1)^2 = q^2$ which are related to
the momenta $p$, $q$, $p_1$, $p_2$, and $p_3$ of the hadrons 
-- $\Omega^*$, $\Xi^*$, $K$, $\Xi$, and $\pi$, 
respectively. The variable $s_1$ has the upper/lower limits $s_1^\pm$ with: 
\eq 
s_1^\pm &=& M_K^2 + M_\Xi^2 
+ \frac{1}{2 s} (s - s_2 - M_K^2) (s_2 + M_\Xi^2 - M_\pi^2) 
\nonumber\\
&\pm& \frac{1}{2 s_2} \, \lambda^{1/2}(s,s_2,M_K^2) \, 
\lambda^{1/2}(s_2,M_\Xi^2,M_\pi^2)\,, \label{threebody_dec} 
\en 
where $\lambda(x,y,z) = x^2 + y^2 + z^2 - 2xy - 2yz -2xz$. 
$\bar u_\Xi(q)$ and $u_{\Omega^*}(p)$ are the spinors corresponding 
to $\Xi$ and $\Omega^*$ baryons, respectively. 
$\Phi(-[p_1 w_{\Xi^*} -  q w_K]^2)$ 
is the correlation function from the Lagrangian~(\ref{LOmega_int}), which 
can be reduced to 
\eq 
\Phi(-[p_1 w_{\Xi^*} -  q w_K]^2) = \exp\biggl[ \frac{M_{\Xi^*}M_K}{\Lambda^2} 
\, \biggl( 1 - \frac{M_{\Omega^*}^2}{(M_{\Xi^*}+M_K)^2}\biggr)\biggr] \,. 
\en  
The factor $c_\pi = \sqrt{2}$ occurs for modes with charged pions $\pi^\pm$ and 
$c_\pi = 1$ for modes with the neutral pion $\pi^0$. $S_{\Xi^*}^{\mu\nu}(q)$ is 
the $\Omega^*$ (fermion spin-$\frac{3}{2}$) propagator introduced in 
Eq.~(\ref{SOmegastar}). 
Note, in our calculations of the three-body decays of the $\Omega(2012)$ 
we use the kinematical formula~(\ref{threebody_dec}), where the invariant mass 
squared $M^2(\Xi\pi)$ is defined by the Mandelstam variable $s_2$. 
We do not impose the condition 1.49 GeV $< M(\Xi\pi) < $ 1.53 GeV used 
in Ref.~\cite{Jia:2019eav} for an efficient signal selection. 

In Figs.~\ref{fig:res1} and~\ref{fig:res2} we show our predictions 
for the respective charged combinations of two- and three-body decay widths
of the $\Omega^*(2012)$ and their sums. 
Results are indicated for the values of $\Lambda = 1$ and 1.5 GeV, 
respectivley, and for the mixing angle $\theta$ varied from 0 to 90 grad.   
In Fig.~\ref{fig:res3} we plot our predictions for 
the ratios $R_1$ and $R_2$ of Eq.~(\ref{R12}) 
of the three- and two-body decays 
and compare them with the upper limits derived by the Belle collaboration 
in Ref.~\cite{Jia:2019eav}. One can see that an increase in the admixture of 
the hadronic component $\Omega\eta$ generates the  suppression of 
the three-body decay widths. For the mixing angle 
bigger $70^\circ$ ($73^\circ$) for $\Lambda = 1$ GeV and 
bigger $55^\circ$ ($61^\circ$) for $\Lambda = 1.5$ GeV 
our ratios are consistent with Belle results for $R_1$ and $R_2$. 
To get a handle on the parameters $\theta$ 
and $\Lambda$ further data on the $\Omega^*(2012)$ decays are needed.
The predictions presented here are strongly correlated with the assumption
that the $\Omega^*(2012)$ has a hadronic molecular structure. 

Finally, we discuss our main numerical results. Varying the model 
scale parameter $\Lambda$ from 1 to 1.5 GeV we find that the magnitude 
of the two-body decay rates slightly increase, while the three-body 
rates decrease. The relative contribution of two to three-body decays 
is governed by the mixing angle $\theta$ --- 
mixing of the $\Xi^* \bar K$ and $\Omega\eta$ 
hadronic components. An increase of $\theta$ generates 
the suppression of the three-body decay widths in comparison with the 
two-body ones. For the total two- and three-body decays rates we get 
the following numerical results: 
\eq 
\Gamma(\Omega^* \to \Xi \bar K) = 2.9 \pm 1.6 \ {\rm MeV}\,, 
\quad  
\Gamma(\Omega^* \to \Xi \pi \bar K) = 1.5 \pm 1.5 \ {\rm MeV} 
\en  
for $\Lambda = 1$ GeV and $\theta \in [0,\pi/2]$, 
\eq 
\Gamma(\Omega^* \to \Xi \bar K) = 6 \pm 3.4 \ {\rm MeV}\,,  
\quad  
\Gamma(\Omega^* \to \Xi \pi \bar K) = 1.4 \pm 1.4 \ {\rm MeV} 
\en  
for $\Lambda = 1.5$ GeV and $\theta \in [0,\pi/2]$. 
The errors in Eqs.(21) and (22) are mainly due to the variation of the mixing
parameter $\theta$. More data on the $\Omega^*(2012)$ the needed to further 
constrain of the parameter $\theta$ and to reduce the errors in our predictions.   
The upper limits of Eq.~(\ref{R12}) set by the Belle  
collaboration~\cite{Jia:2019eav} for the ratios of   
three- and two-body decay rates of the $\Omega(2012)$ 
are fulfilled for following values of the $\theta$ angle:
 
1) $\Lambda = 1$ GeV 
\eq 
R_1 < 9.3 \% \ \ \ {\rm at} \ \ \theta \ge 70^\circ \,, \quad\quad 
R_2 < 7.8 \% \ \ \ {\rm at} \ \ \theta \ge 73^\circ \,. 
\en 

2) $\Lambda = 1.5$ GeV 
\eq 
R_1 < 9.3 \% \ \ \ {\rm at} \ \ \theta \ge 55^\circ \,, \quad\quad 
R_2 < 7.8 \% \ \ \ {\rm at} \ \ \theta \ge 61^\circ \,. 
\en  
We also estimated limits for the ratio of
three- to two-body modes 
\eq
R   = \frac{B(\Omega(2012) \to \Xi(1530)(\to \Xi\pi)\bar K)}
           {B(\Omega(2012) \to \Xi \bar K)} \,. 
\en 
For $\theta \ge 81.3^\circ$ at $\Lambda = 1$ GeV and
    $\theta \ge 72.8^\circ$ at $\Lambda = 1.5$ GeV 
we fulfill the limits of the Belle collaborations $R < 11.9 \%$. 
Hence a sizable $\Omega\eta$ hadronic component in the $\Omega^*(2012)$ 
leads to a suppression of the $\Xi\pi \bar K$ mode relative to the 
$\Xi \bar K$ 
channel. To get a further handle on the parameters $\theta$ and $\Lambda$ 
in the context of the hadronic structure precise data on the $\Omega^*(2012)$ 
decays are needed. The prediction given here can hopefully support a possible 
structure interpretation of the $\Omega^*(2012)$.  
                                                          
\section*{Acknowledgments}

This work was funded by Bundesministerium f\"ur Bildung und Forschung
``Verbundprojekt 05P2018 - Ausbau von ALICE 
am LHC: Jets und partonische Struktur von Kernen''
(F\"orderkennzeichen No. 05P18VTCA1),
``Verbundprojekt 05A2017-CRESST-XENON'' 
(F\"orderkennzeichen 05A17VTA)'', 
the Carl Zeiss Foundation (Project Gz: 0653-2.8/581/2),  
ANID PIA/APOYO AFB180002 (Chile),  FONDECYT (Chile) Grant No. 1191103,
Tomsk State and Tomsk Polytechnic University Competitiveness Enhancement 
Programs (Russia).

\end{document}